\documentclass[aps, pra, reprint, amsmath, amssymb, showpacs]{revtex4-1}
\usepackage{graphicx}
\usepackage{dcolumn}
\usepackage{bm}
\usepackage{amsmath}
\usepackage{comment}
\usepackage[dvipsnames]{xcolor}
\DeclareMathOperator*{\SumInt}{%
\mathchoice%
  {\ooalign{$\displaystyle\sum$\cr\hidewidth$\displaystyle\int$\hidewidth\cr}}
  {\ooalign{\raisebox{.14\height}{\scalebox{.7}{$\textstyle\sum$}}\cr\hidewidth$\textstyle\int$\hidewidth\cr}}
  {\ooalign{\raisebox{.2\height}{\scalebox{.6}{$\scriptstyle\sum$}}\cr$\scriptstyle\int$\cr}}
  {\ooalign{\raisebox{.2\height}{\scalebox{.6}{$\scriptstyle\sum$}}\cr$\scriptstyle\int$\cr}}
}

\begin{document}

\title{Two-photon Above Threshold Ionization of Helium}
\author{Yimeng Wang and Chris H. Greene}
\affiliation{Department of Physics and Astronomy, Purdue University, West Lafayette, Indiana 47907, USA }
\affiliation{Purdue Quantum Science and Engineering Institute, Purdue University, West Lafayette, Indiana 47907, USA }

\begin{abstract}
Multiphoton ionization provides a clear window into the nature of electron correlations in the helium atom. In the present study, the final state energy range extends up to the region near the $N=2$ and $N=3$ ionization thresholds, where two-photon ionization proceeds via continuum intermediate states above the lowest threshold. Our calculations are performed using multichannel quantum defect theory (MQDT) and the streamlined R-matrix method. The sum and integration over all intermediate states in the two-photon ionization amplitude is evaluated using the inhomogeneous R-matrix method developed by Robicheaux and Gao. The seamless connection of that method with MQDT allows us to present high resolution spectra of the final state Rydberg resonances.  Our analysis classifies the resonances above the $N=2$ threshold in terms of their group theory quantum numbers.  Their dominant decay channels are found to obey the previously conjectured propensity rule far more weakly for these even parity states than was observed for the odd-parity states relevant to single photon ionization. 
\end{abstract}

\maketitle
\section{introduction}
Single photon absorption has been used since the 1960s to probe doubly-excited states with extreme electron correlation in the helium atom and many other atoms.\cite{macek1968JPB,fano1983RPP,lin1986AMOP}  In recent decades, increasing interest has derived from studying electron correlation in helium using other types of excitation, such as non-sequential multiphoton double ionization which has received extensive attention.\cite{Pindzola:1998,Feng:2003, Nikolopoulos:2001}  Additionally, the use of multiple fields to dress the helium atom Hamiltonian with infrared (IR) light has demonstrated remarkable control possibilities, both for autoionization line shapes and temporal pulse phase control.\cite{OttScience,LinChu2013,GaardeSchafer2013,ChuLin2014}

Multiphoton ionization can in principle proceed either via sequential steps through one or more bound intermediate stationary states en route to ionization, or non-sequentially with either no intermediate stationary state or else through a continuous intermediate state in the ionization continuum.  In the latter case, the multiphoton ionization is often referred to as an above-threshold ionization process, because more photons are absorbed than the minimum number required to ionize the atom.

\textcolor{black}{The present study quantitatively treats the process of two-photon double excitation of helium to the lower-lying autoionizing states, 
where at least one of the electrons is in either the $N = 2$ or $N = 3$ shell. 
This process is very much related to the two-photon double ionization\cite{Pindzola:1998,Feng:2003,Nikolopoulos:2001}, but is  experimentally simpler and theoretically less demanding. 
The photoelectron angular distribution is computed to elucidate the strengths of interfering pathways associated with direct ionization and with correlated autoionization decay. 
Based on a combination of the R-matrix and generalized quantum defect methods, 
 we achieve good consistency with some previous calculations where our results can be compared with them, and extend them into higher energy ranges not routinely treated. 
Then we classify the resonances near the $N = 3$ threshold according to the $SO(4)$ correlated classification scheme and test the propensity rules that have been confirmed to operate in the $^1P^o$ states reached by the single-photon ionization, which have never been discussed in the multi-photon region. }

\textcolor{black}{The structure of this paper is organized as follows: In Sec. \ref{method} we introduce the underlying theory and  methods for our calculations. The setup of our study including the photoionization Hamiltonian is given in subsection \ref{ss1}. An introduction on our numerical treatment is given in subsections \ref{ss2} and \ref{ss3}, and the correlated classification scheme is briefly described in subsection \ref{ss4}. 
Sec. \ref{result} presents our calculation results, subsection \ref{rs1} gives the generalized cross section and the photoelectron angular distribution, while subsection \ref{rs2} discusses the propensity rules according to the classification scheme of the resonances. 
Finally, Sec. \ref{end} summarizes the main conclusions of our research.}

\section{METHODOLOGY}\label{method}

\subsection{Light-atom interaction}\label{ss1}
The nonrelativistic Hamiltonian for a helium atom interacting with a monochromatic laser source is the following, neglecting fine-structure (in atomic units, $m_e=\hbar=e=1$):
\begin{equation}
\label{eq1-1}
\begin{split}
    &H(\vec{r}_1,\vec{r}_2,t)=H_0(\vec{r}_1,\vec{r}_2)+V(\vec{r}_1,t)+V(\vec{r}_2,t)\\
    &H_0(\vec{r}_1,\vec{r}_2)=\sum_{i=1,2}\left( -\frac{\nabla_{\vec{r}_i}^2}{2} -\frac{2}{r_i}\right)+\frac{1}{|\vec{r}_1-\vec{r}_2|}\\
    &V(\vec{r}_i,t)=\frac{1}{2}\mathcal{E}_0\hat{\mathbf{\epsilon}} \cdot \vec{r}_i e^{-i\omega t}\qquad (i=1,2)
\end{split}
\end{equation}

Here $H_0$ is the non-relativistic helium Hamiltonian whose nucleus is fixed at origin and treated as infinitely massive.
$V$ is the interaction between an electron and the laser field in the electric dipole approximation. $\mathcal{E}_0$ is the electric field amplitude, $\omega$ is the angular frequency, and $\hat{\epsilon}$ is the polarization unit vector. Since only absorption processes are considered, the $e^{+i\omega t}$ term is omitted in the usual rotating wave approximation. For simplicity only a single laser source is included in Eq.\eqref{eq1-1}. 

The generalized cross section (sometimes abbreviated in the text by ``cross section'') for two-photon ionization is the following: 
\begin{equation}
\label{eq2-3}
    \sigma_{tot}=2\pi(2\pi\alpha\omega)^2\sum_{f} \left|T_{0,f}\right|^2 .
\end{equation}
\textcolor{black}{
Here $\alpha$ is the fine structure constant, $0$ and $f$ denote all the quantum numbers for initial and final states,
 $T_{0,f}$ is the transition amplitude from the initial to the final states. Based on a second order perturbation treatment of the light-atom interaction, which is accurate in the low intensity limit for a monochromatic laser source, the formula for $T_{0,f}$ is: 
}
\begin{equation}
\label{eq2-1}
\begin{split}
     &T_{0,f} =\SumInt_{\zeta} dE_{\zeta}\frac{\langle f|D| {\zeta} \rangle
    \langle {\zeta} |D| 0\rangle}{E_0+\omega-E_{\zeta}}  \\
    &=\SumInt_{\zeta} dE_{\zeta}\frac{\langle f||D|| {\zeta} \rangle
    \langle {\zeta} ||D|| 0\rangle}{E_0+\omega-E_{\zeta}}(-1)^{L_f + L_{\zeta}-M_f-M_{\zeta}}\\
    &\times\sum_{qq^{\prime}} (-1)^{q+q^{\prime}}\epsilon_{-q}^{(1)}\epsilon_{-q^{\prime}}^{(1)}   \left(
    \begin{matrix}
    L_f & 1& L_{\zeta} \\
    -M_f & q^{\prime} & M_{\zeta}
    \end{matrix}
    \right)
        \left(
    \begin{matrix}
    L_{\zeta} & 1& L_0 \\
    -M_{\zeta}& q & M_0
    \end{matrix}
    \right)    
\end{split}
\end{equation}

\textcolor{black}{
Here we give our notations for the wavefunctions in this paper: $\Psi_{f}(\vec{r}_1,\vec{r}_2)=\langle\vec{r}_1,\vec{r}_2|f\rangle$ is the final state wavefunction of bare helium atom $H_0$, with energy, total angular momentum, and projected angular momentum be $E_f$, $L_f$, and $M_f$. Substituting the subscript from $f$ to $0$ and $\zeta$ gives same things for the initial and intermediate states. Because the total spin $S$ is a conserved quantum number for the nonrelativistic spin-independent Hamiltonian used in the present study, we do not explicitly write the spin-degrees of freedom in our notation.
$D= \hat{\mathbf{\epsilon}} \cdot (\vec{r}_1+\vec{r}_2)$ is introduced to denote the 
dipole operator in the length gauge. 
The intermediate energy $E_{\zeta}$ includes all eigenvalues of $H_0$ that obey the usual good symmetry and angular momentum selection rules. The notation $\SumInt$ is used to summed over intermediate states, since $|{\zeta}\rangle$ can represent both bound and continuous eigenstates.
The continuous intermediate and final states are normalized per unit energy, and the final state is subjected to the usual incoming wave boundary condition asymptotically. }

\textcolor{black}{The second equality in Eq. \eqref{eq2-1} is written in terms of reduced matrix elements, giving all the projected angular momentum orientations explicitly according to the Wigner-Eckart theorem. The bracket () denotes the Wigner-3J symbol. $\epsilon_{q}^{(1)}$ are the spherical components of the rank-1 polarization tensor, 
$\epsilon_{0}^{(1)}=\hat{z}$ and $\epsilon_{\pm1}^{(1)}=\mp1/\sqrt{2}(\hat{x}\pm i\hat{y})$.}

The angular distribution of the photoelectron, i.e., the differential generalized two-photon ionization cross-section, is given by:  
\begin{scriptsize}
\begin{equation}
\label{eq2-4}
\begin{split}
\frac{d\sigma}{d\Omega}&=8\pi^3\alpha^2\omega^2\sum_{n_1l_1m_1} \left|\sum_{l_2m_2L_fM_f} Y_{l_2m_2}(\hat{k})\langle l_1m_1,l_2m_2|L_f,M_f\rangle T_{0,f}\right|^2 \\
&=\frac{\sigma_{tot}}{4\pi}\sum_{j} \beta_j(\omega)P_j(cos\theta)
\end{split}
\end{equation}
\end{scriptsize}
\textcolor{black}{
The subscript $1(2)$ denotes for the single-electron quantum numbers for the core(emission) electron. The second equality follows only when $M_f=0$, which is the situation will be discussed in the present study. }
The differential cross-section can be written in terms of Legendre polynomials $P_j(cos\theta)$ and asymmetry parameters $\beta_j$, 
 where $\theta$ is the polar angle between the ejected electron direction and the polarization axis.
Our normalization choice constrains $\beta_0$ to equal 1. Therefore, $\beta_j\neq0$ only for $j=0,2,4$ for two-photon ionization, with explicit expressions for $\beta_{2,4}$ given by:
\begin{equation}
\label{eq2-2}
\begin{split}
   &\beta_j^{(n_1)} = \frac{1}{\sigma_{(n_1)}}\sum_{l_1}\sum_{l_2,l_2^{\prime},L_f,L_f^{\prime}} (-1)^{l_1+l_2+l_2^{\prime}+L_f+L_f^{\prime}}T_{0,f}T_{0,f^{\prime}}^{*}\\
   &\left[l_2\right]\left[l_2^{\prime}\right]\left[L_f\right]\left[L_f^{\prime}\right]\left[j\right]^2
     \left(
    \begin{matrix}
    l_2 & l_2^{\prime}& j \\
    0 & 0 & 0
    \end{matrix}\right)\left(
    \begin{matrix}
    j & L_f & L_f^{\prime} \\
    0 & 0 & 0
    \end{matrix}\right)\left\{
    \begin{matrix}
    j & L_f & L_f^{\prime} \\
    l_1 & l_2^{\prime} & l_2
    \end{matrix}\right\} \\
    &\beta_j= \sum_{n_1}\frac{\sigma_{(n_1)}}{\sigma_{tot}}\beta_j^{(n_1)}\\
\end{split}
\end{equation}
\textcolor{black}{Of interest are the separate angular distribution asymmetry parameters associated with each separate group of photoelectrons with different energies, denoted in the present notation as $\beta_j^{(n_1)}$. These correspond to electrons with a certain kinetic energy escape in all channels having a given ionic principal quantum number $N = n_1$. $\sigma_{(n_1)}$ is the corresponding partial cross section for final states with channel $N = n_1$, which is introduced to ensure $\beta_0^{(n_1)} \equiv 1$. 
The bracket $\left\{ \right\}$ represents the Wigner-6J symbol, and $[L]\equiv\sqrt{2L+1}$.}
The derivation details of \eqref{eq2-2} largely follow the discussion in \cite{Lindsay:1992}.

\subsection{R-matrix MQDT method}\label{ss2}

One of the main issues in this problem is to determine the eigenstates for a bare nonrelativistic helium atom $H_0(\vec{r}_1,\vec{r}_2)$. 
This is a quantum-three-body problem and our theoretical treatment includes all the electron correlation in the energy range chosen for exploration. 
The energy eigenstates of this system are calculated using the streamlined eigenchannel R-matrix method \cite{Review:1996} and multi-channel quantum defect theory (MQDT). \cite{Review:1996,Greene:1985,Seaton:1983}. \textcolor{black}{The following is a brief introduction to the MQDT method, which can also be found in the literature cited above.} 

The idea is to choose a reaction volume or box in the independent electron radii with boundaries $R_0$ for both electron distances from the nucleus, and tackle the problem separately inside and outside. Because the energies considered here are well below the double-ionization threshold, our solution neglects the possibility that both electrons escape from the box simultaneously.  Methods for handling two-electron escape within the present framework have been considered in previous treatments.\cite{MeyerGreene1994pra,MeyerGreeneEsry1997prl}  The correlated motion of electrons is fully considered only when both the electrons are inside the box, although the dipole interaction that operates when one electron is outside the box implies long-range correlations that are described through the Gailitis-Damburg transformation of asymptotic channels.\cite{GailitisDamburg,Greene:1979} 
In the streamlined R-matrix method, the wavefunctions inside the box are expanded into antisymmetrized single-particle product basis functions, with the expansion coefficients determined by the variational method. 
In the regions where one electron escapes beyond the boundary (e.g. $r_2>R_0$), its interaction with the residual ion can be approximated by the lowest one or two terms in the multipole expansion of $\frac{1}{r_{12}}$. 
At the simplest level of approximation $H_0$ can be treated by the following form that approximates $\frac{1}{r_{12}} \approx \frac{1}{r_{2}} $, which has separable eigensolutions, 

\begin{equation}
    H_0(\vec{r}_1,\vec{r}_2)\stackrel{r_2>R_0}{\longrightarrow}\left(-\frac{\nabla_{\vec{r}_1}^2}{2}-\frac{2}{r_1}\right)+\left(-\frac{\nabla_{\vec{r}_2}^2}{2}-\frac{1}{r_2}\right).
\end{equation}
At this level of approximation, only the screened Coulomb interaction is experienced by the outer electron and the long range dipole interaction term $\vec{r}_1\cdot\hat{r}_2/r_2^2$ is neglected. 
In this representation, the quantum numbers $(n_1,l_1;\epsilon_2,l_2)$ can be used to describe the electron escape channels, in addition to the good quantum numbers of the unperturbed atom, namely $\{L,S,\pi\}$.
The $i$-th linearly independent eigenfunction at energy $E$ can then be written as follows in the region $r_1<R_0$ and $r_2>R_0$: 
\begin{equation}
\Psi_{i}(\vec{r}_1,\vec{r}_2)=\mathcal{A}\left(r_2^{-1}\sum_{j=1}^{n}\Phi_j(\vec{r}_1,\Omega_2)\psi_{ji}(r_2)\right)
\end{equation}
The channel functions $\Phi_j(\vec{r}_1,\Omega_2)$ contain the inner electron radial wavefunction $u_{n_1,l_1}(r_1)/r_1$ multiplied by coupled spherical harmonics $Y_{l_1,l_2,LM}(\Omega_1,\Omega_2)$ and the total spin wavefunction of the system;  these channels for a given set of $\{L,S,\pi\}$ are characterized by the channel quantum numbers $j\equiv(n_1,l_1,l_2)$. They are associated with ionization threshold energies $\epsilon^{thresh}$ which are the inner electron energy levels $\epsilon^{thresh}_j=\epsilon^{thresh}_{n_1,l_1}$.  $\psi_{ji}(r_2)/r_2$ describes the radial motion of the outer electron whose channel energy outside the box is $\epsilon_j$. \textcolor{black}{$\mathcal{A}$ is the antisymmetrizer that exchanges the status of electrons $1$ and $2$.}
$j$ is the channel index and $i$ labels the degenerate, linearly-independent solutions. 
Depending on the sign of $\epsilon_j=E-\epsilon^{thresh}_j$, the channels can be separately grouped into open and closed subsets.
The open channels ($\epsilon_j>0$) give the available continuum electron states that photoelectrons can escape into, all the way to infinity.
The closed channels ($\epsilon_j<0$) describe channels that support weakly bound doubly-excited atomic states that are quasi-bound resonance states.

The connection information for the solutions across the box boundary at $R_0$ is given by the R-matrix, $\underline{R}= \underline{\psi} \left[\underline{\psi}^\prime\right]^{-1}$,  which is the inverse of the logarithmic derivative matrix of the outermost electron radial wavefunction at $R_0$.
The $j-th$ component of wavefunction in the outer region, $r_2=R_0$ to $\infty$, is written as a linear combination of the two radial reference solutions in the long range potential for the $j$-th channel, $\hat{f}_j(r_2), \hat{g}_j(r_2)$ \cite{Greene:1979,Seaton:1983} which are viewed in this notation as diagonal matrices.  The wavefunction outside the box can be written as: $\underline{\psi}(r_2)=\underline{\hat{f}}(r_2)-\underline{\hat{g}}(r_2)\underline{K}^{sr}$,
where $\underline{\psi}(r_2)$ is matched to the R-matrix by choosing 
$\underline{K}_{sr}=(\underline{\hat{f}}(R_0)-\underline{\hat{f}}^\prime(R_0)\underline{R})(\underline{\hat{g}}(R_0)-\underline{\hat{g}}^\prime(R_0)\underline{R})^{-1} $.
In the closed channels at $r_2\rightarrow\infty$, exponential growth of $\underline{\psi}(r_2)$ must be eliminated in the sense of multichannel quantum defect theory(MQDT) by finding an appropriate linear transformation $\underline{Z}$. The standard formulas of MQDT show that the exponentially growing terms can be eliminated by a transformation shown, e.g., in Ref.\cite{Review:1996} to have the form: 
$\underline{\psi}^{lr}(r_2)=\underline{\psi}(r_2)\underline{Z}$. Finding this matrix $\underline{Z}$ is accomplished using the traditional formulas of MQDT.

The other major issue is to evaluate the sum over all the intermediate states in Eq.\eqref{eq2-1}.
The inhomogeneous R-matrix method is implemented as was formulated by Robicheaux and Gao \cite{Robicheaux:1991,Robicheaux:1993} to solve this problem. 
The idea is to replace the infinite sum by an integral over the inhomogeneity multiplied by the Green’s function, as in the Dalgarno-Lewis method.
Instead of calculating the sum and integral implied by  \textcolor{black}{ the expression $|\Lambda_p\rangle=\int dE_{\zeta} |{\zeta} \rangle \frac{\langle {\zeta} |D| 0\rangle}{E_0+\omega-E_{\zeta}}$, this method solves the inhomogeneous equation, $(E_0+\omega-H_0)\Psi_{\Lambda_p}= D \Psi_0$. The function $\Psi_{\Lambda_p}$ can be viewed as the intermediate virtual state after the atom absorbs the first photon.}
Again, the R-matrix plus MQDT method is utilized to solve for $\Psi_{\Lambda_p}$, but with different boundary conditions imposed. 
Rather than being a stationary standing wave energy eigenstate as is the case for $\Psi_0$ and $\Psi_f$, this is asymptotically a purely outgoing wave in the open channels, and exponentially decaying in the channels that are closed at the intermediate state energy (i.e., the ground state energy plus one photon energy). 
\textcolor{black}{Therefore the streamlined R-matrix calculation for this inhomogeneous equation takes a rather different form from the R-matrix calculation used to solve the  homogeneous equation $(E-H_0)\Psi= 0$. \cite{Robicheaux:1993}.}

\subsection{Gailitis-Damburg transformation}\label{ss3}
In the MQDT treatment discussed in the previous subsection, the potential experienced by the outer electron, 
$-\frac{2}{r_2}+\frac{1}{|\Vec{r}_1-\Vec{r}_2|}$, was approximated by the overall Coulomb interaction with the screened He$^+$ ion, namely $-\frac{1}{r_2}$, which neglects the electron correlations outside the R-matrix box. Now we consider the strongest electron correlations associated with coupling between the outer electron and the permanent electric dipole moment of the degenerate excited He$^+$ core, i.e. $\frac{1}{|\Vec{r}_1-\Vec{r}_2|}\approx \frac{1}{r_2}+\frac{\vec{r_1}\cdot\hat{r_2}}{r_2^2}$.
Owing to the degeneracy of the excited hydrogenic thresholds (neglecting fine structure and the Lamb shift), this dipole term can be absorbed into the threshold Hamiltonian and merged with the outer electron angular momentum term $\frac{\hat{l}_2^2}{2r_2^2}$. 
We denote the composite of the dipole and the orbital (centripetal) term as the Gailitis-Damburg operator\cite{GailitisDamburg}, whose matrix elements are written as

\begin{equation}
\begin{split}
    A_{ij}&=\langle \phi_i(r_1)Y_{l_1^i,l_2^i,L}|\hat{l}_2^2+2r_1\cos{\theta_{12}}| \phi_j(r_1)Y_{l_1^j,l_2^j,L}\rangle    \\
     &=l_2(l_2+1)\delta_{l_2^i,l_2^j}+2\langle\phi_i|r_1|\phi_j\rangle \\
     &\times\langle (l_1^i,l_2^i)L|\cos{\theta_{12}}|(l_1^j,l_2^j)L\rangle .\\
\end{split}
\end{equation}
\textcolor{black}{Here $\phi_i(r_1)Y_{l_1^i,l_2^i,L}$ is the explicit form of the spatial part of the channel function $\Phi_i(\vec{r}_1,\Omega_2)$.}
The Gailitis-Damburg matrix can be represented in terms of its eigenvectors and eigenvalues as $A_{ij}=\sum_{\gamma}X_{i,\gamma}a_{\gamma}X^T_{\gamma,j}$,
which then allows us to perform a generalized MQDT solution of the outer region radial equation in the dipole representation with the new channel index $\gamma$. The radial Schrodinger equation for the outer electron in the channel $\gamma$ is,
\begin{equation}
    (-\frac{1}{2}\frac{d^2}{dr_2^2}-\frac{1}{r_2} +\frac{a_{\gamma}}{2r_2^2})\mathcal{F}_f^{\gamma}(r_2)= \epsilon_{\gamma} \mathcal{F}_f^{\gamma}(r_2)
\end{equation}

\textcolor{black}{The channel correspondence between the independent electron basis $\{i\}=(n_1,l_1;l_2)$ and the Gailitis-Damburg basis $\{\gamma\}$ is controlled by the eigenvector matrix $X_{i,\gamma}$.\cite{SGC1992} }
We can define a real or complex angular momentum $\tilde{l}_{\gamma}$ by considering rewriting the $\gamma$-th eigenvalue of $A$ as $a_{\gamma}\equiv\tilde{l}_{\gamma}(\tilde{l}_{\gamma}+1)$. For the repulsive dipole case where $a_\gamma > -1/4$, $\tilde{l}_{\gamma}$ is real, and for attractive dipole case $\tilde{l}_{\gamma}$ is complex.  The details for the new reference wavefunction $\mathcal{F}_f^{\gamma}(r_2)$ and its corresponding long-range QDT parameters can be found in \cite{Greene:1979,GRF1982, ErratumGRF1982}, as are the regular and irregular radial solutions in the dipole potentials.

\begin{figure}[htbp]
  \includegraphics[scale=0.28]{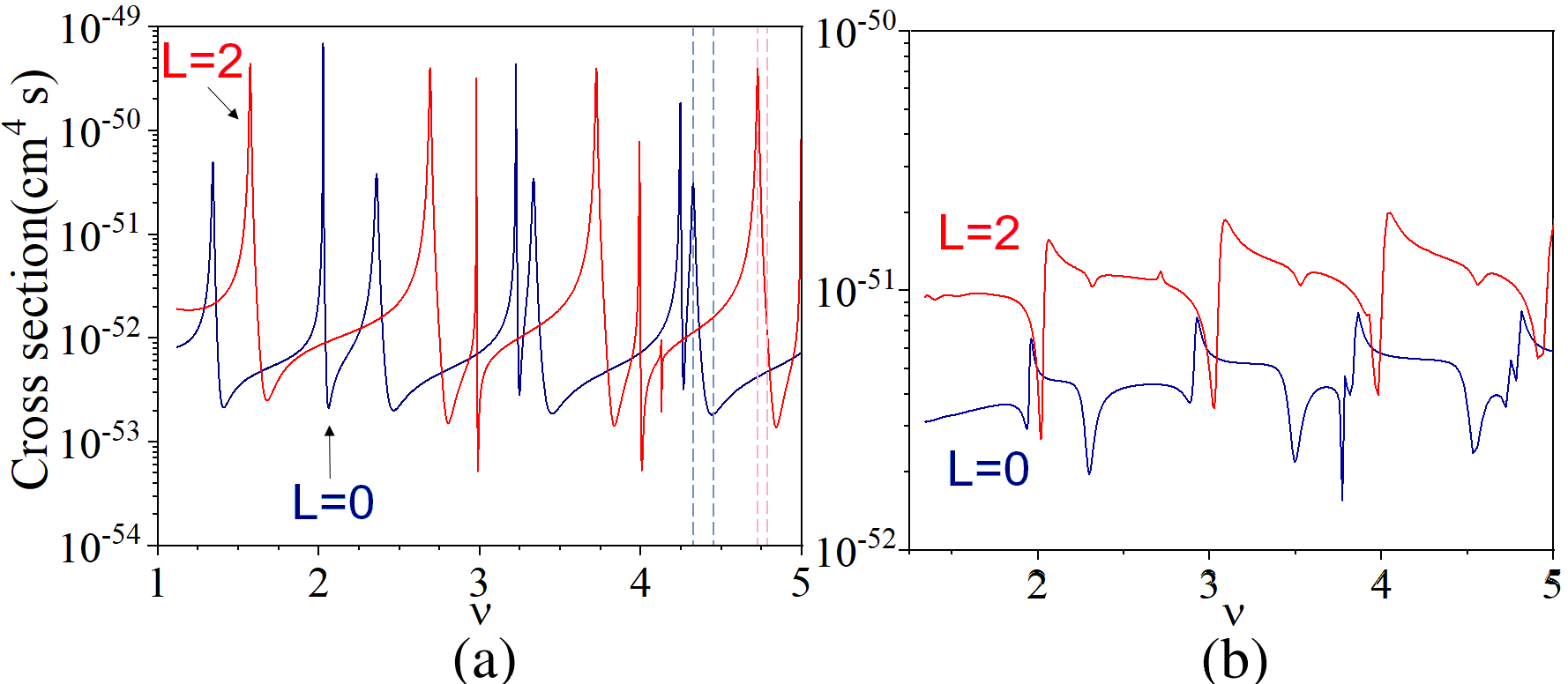}
    \caption{ \textcolor{black}{The calculated total generalized two-photon ionization cross sections, with inclusion of the long range Gailitis-Damburg dipole interaction term included outside the R-matrix box, are presented in the vicinity of the first two excited ionization thresholds. The left panel is calculated at energies below the $N=2$ threshold and the right panel at energies below the $N=3$ threshold.} The R-matrix box radii for the two calculations are at 15 a.u. and 34 a.u. respectively.  \textcolor{black}{With the Gailitis-Damburg dipole term included, the position of resonances are shifted by around $\Delta\nu \approx 0.1$ (the level shift is indicated by the dashed vertical lines), relative to calculations that neglect that long-range charge-dipole physics.} }
  \label{11}
\end{figure}

The Gailitis-Damburg treatment influences our results, as can be deduced from the two-photon cross-sections plotted up to the He$^+$ $N=2$ and $N=3$ thresholds in Fig.\ref{11}. The horizontal axes are shown on an effective quantum number scale $\nu=1/\sqrt{2(E_{N}-E_f)}$. 
At small $\nu$, there are no differences since when both electrons are in a resonance that is deeply bound, their motions are almost restricted inside the box(at $R_0=15(34) a.u.$ for $N=2(3)$) and are determined by the R-matrix calculations that incorporate all the electron correlations. \textcolor{black}{The differences appear only when the effective quantum number is high enough that the outer electron is active outside the box. In this regime, the Gailitis-Damburg-type treatment shifts the positions of resonances by approximately $\Delta\nu \approx 0.1$ (this shift is indicated by the dashed vertical lines on the figure).}
The results using the Gailitis-Damburg channel representation outside of the R-matrix box include more physics and are therefore more reliable; therefore, in the remainder of the paper all of our results are based on the Gailitis-Damburg long-range quantum defect theory treatment.

For the present study, a variational basis of radial B-splines are implemented to construct the single-particle basis functions. For final state energies up to $N=2$ threshold, the R-matrix boundary $R_0$ is set to be 15 a.u., the number of radial part basis functions is 20 per $l$-value, and single particle angular momenta are included in the range $l=0-4$. In the R-matrix calculation, the total number of basis functions are 855, 1296, and 1656 closed-type two-electron bases for the singlet $L = 0$, $L = 1$, and $L = 2$ symmetries. 
For energies up to the $N=3$ threshold, $R_0$ is taken to be 34 a.u., with 40 radial basis functions per $l$, and with single particle angular momenta included in the range $l=0-9$. The number of closed-type configurations for $L = 0$, $L = 1$, and $L = 2$ states are, 3990, 6787 and 9286 respectively.

\subsection{The classification scheme}\label{ss4}
This subsection introduces the classification of resonances in the $_N(K,T)^A_n$  scheme following Herrick and Sinanoglu\cite{HerrickSinanoglu} and Lin\cite{Lin1984}, who introduced these correlation quantum numbers to describe the strong electron-electron correlation in two-electron atoms. $N$ and $n$ are the principal quantum numbers for the inner and outer electrons ($n_1$ and $n_2$). The $A$ quantum number describes the radial motion correlation, and can take values +1, 0, and -1 only. $A=+1(-1)$ means there is an approximate antinodal(nodal) structure at or near the line $r_1=r_2$, and $A=0$ means there is no such structure. 
$K, T$ describe the angular correlations, which have their origin in a group theory analysis\cite{HerrickSinanoglu}. A single electron bounded in the Coulomb field can be fully determined by its angular momentum $\vec{l}$ and Runge-Lenz vector $\vec{a}$. 
To analyse a two-electron system, we consider the coupling vectors $\vec{L}=\vec{l}_1+\vec{l}_2$ and $\vec{B}=\vec{a}_1+\vec{a}_2$, that remain unchanged under the first order of electron interaction\cite{Rau:1990}.
For given values of $L$ and parity $\pi$, the quantum numbers $K, T$ determine the $O_4$ group Casimir invariants $\vec{B}^2$ and $\vec{B}\cdot\vec{L}$. 
$T$ has been identified with the pseudoscalar operator $(\vec{L}\cdot\vec{r}_{12})$, and $K$ is proportional to $-\langle\vec{r}_1\cdot\hat{r}_2\rangle$. 
The possible values of $K$ and $T$ for given $N$ and $L$ are given by $T=O, 1,2,...,min(L,N- 1)$ and $K=N-1-T,N-3-T, ...,-(N-1-T)$. The maximum positive value of $K$ has the most attractive dipole term that decreases the long-range potential. 

\begin{figure}[htbp]
  \includegraphics[scale=0.29]{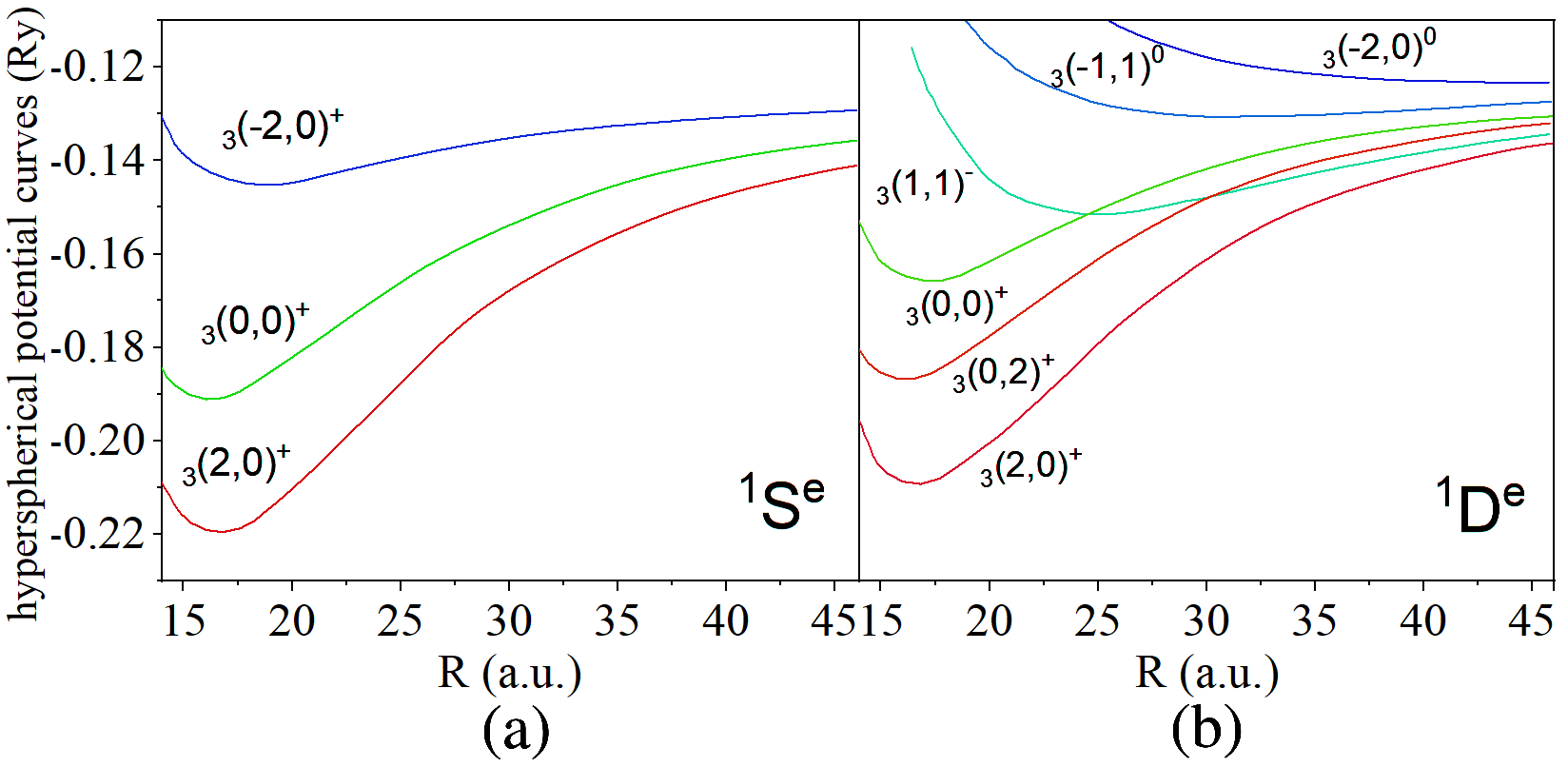}
  \caption{Hyperspherical potential curves for $^1S^e$ and $^1D^e$ for the helium atom that converge to the He$^+$ N=3 threshold. Reduced units with $Z=1$ are used, as defined by Lin, namely the horizontal axis has units of Bohr radius divided by $Z=2$ while the vertical (energy) axis has units of Hartree multiplied by $Z^2$.
  Reprinted from Fig.1. of Lin.\cite{Lin1984}. }
  \label{13}
\end{figure}

In hyperspherical coordinates, on the other hand, the approximate wave functions in the adiabatic approximation are expressed as $F_i(R)\Phi_i(R,\Omega)$, where $\Omega=(\alpha,\theta_{12})$, $R=\sqrt{r_1^2+r_2^2}$ and $\alpha=tan^{-1}(r_2/r_1)$. $R$ specifies the overall size of the atom and does not describe electron correlation effects; rather, all the electron-electron correlation is included in the channel functions $\Phi_i(R,\Omega)$, at least in the adiabatic approximation which works reasonably well here.\cite{lin1986AMOP,Lin1984,Sadeghpour1991} The channel potential curves for $^1S^e$ and $^1D^e$ wave are shown in Fig.\ref{13}. 
At small R, the $A=+$ curves are the lowest ones whose potential curve minima occur at the smallest values of $R$, so the electrons in these channels spend considerable time near the nucleus where they interact strongly with each other, making their autoionization states  unstable and decay relatively rapidly; the corresponding resonances are relatively broader than states in the other nearby channels. At large R, the ordering of the potential curves is determined by the $K$ values, as the differences of the long-range potentials are controlled by the dipole interaction. 
It is possible to ascertain the connection between the $_N(K,T)^A$ channels and the Gailitis-Damburg channels at large distance, as is explained in the following paragraph. 

Our R-matrix box radius is set at $R_0=34$, outside of which there are no more crossings among the hyperspherical curves. With the aid of Fig.\ref{13}, we can deduce the correspondence between the $_N(K,T)^A$ channels and the Gailitis-Damburg channels $\gamma$ by matching the largest $a_{\gamma}$ to $(K,T)=(-2,0)^+$ for $^1S^e$ symmetric or to $(K,T)=(-2,0)^0$ for $^1D^e$ symmetric and so on. 
This analysis allows us to classify the resonances in the $_N(K,T)^A$ scheme of approximately good quantum numbers and to discuss the validity of propensity rules for photoexcitation and decay.

\section{results and discussion}\label{result}
\subsection{Two-photon ionization and angular distributions}\label{rs1}
\begin{figure}[htbp]
  \includegraphics[scale=0.32]{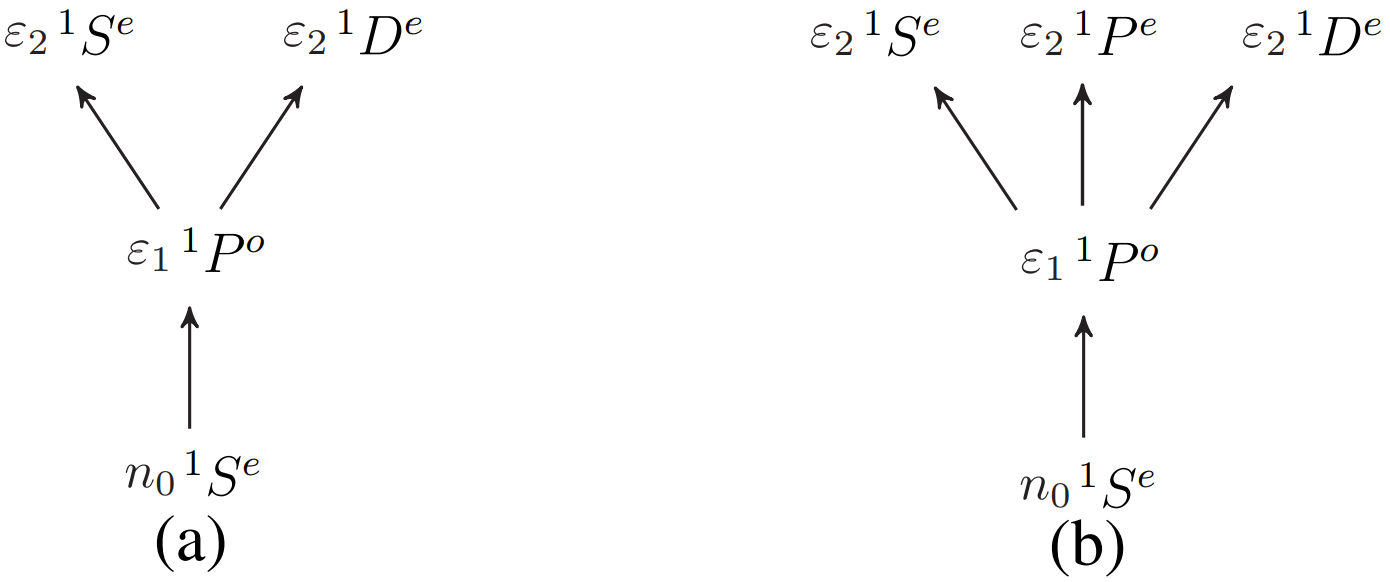}
  \caption{\textcolor{black}{Schematic diagram of (a) one- and two-photon ionization channels allowed by parallel linearly polarized electric-dipole selection rules. (b) one- and two-photon ionization channels allowed by two opposite circularly polarized fields. In both the two diagram, $^{2S+1}L^{\pi}$ denotes a doubly-excited helium autoionizing state with total orbital angular momentum $L$, total spin state $S$, and  even(+1) or odd(-1) parity $\pi$.  The helium atom spins are assumed to remain in a singlet state. The $^1P^e$ state is a parity-unfavored state whose lowest channel becomes open at the $N=2$ hydrogenic threshold.}}
  \label{1}
\end{figure}
\textcolor{black}{
This section presents the cross-section and angular distributions for two-photon ionization of ground-state helium ($1s^2$, singlet). 
To start with, we give the possible values of the quantum numbers $^{2S+1}L^{\pi}$ in Fig.~\ref{1}, with $L$ and $S$ the quantum number for total angular momentum $\vec{L}=\vec{l}_1+\vec{l}_2$ and spin momentum $\vec{S}=\vec{s}_1+\vec{s}_2$, and $\pi$ the parity of helium, which is determined by individual orbital momenta as $\pi=(-1)^{l_1+l_2}$. Fig.~\ref{1}(a) shows the situation for parallel linearly polarized photons and (b) for opposite circularly polarized photons.
Since the helium atom is accurately described by the LS-coupling scheme, the atom-light interaction can only change $L$, whereas $S$ remains invariant, that only singlet spin state will show up in our calculation.}

\begin{figure*}[htbp]
  \includegraphics[scale=0.39]{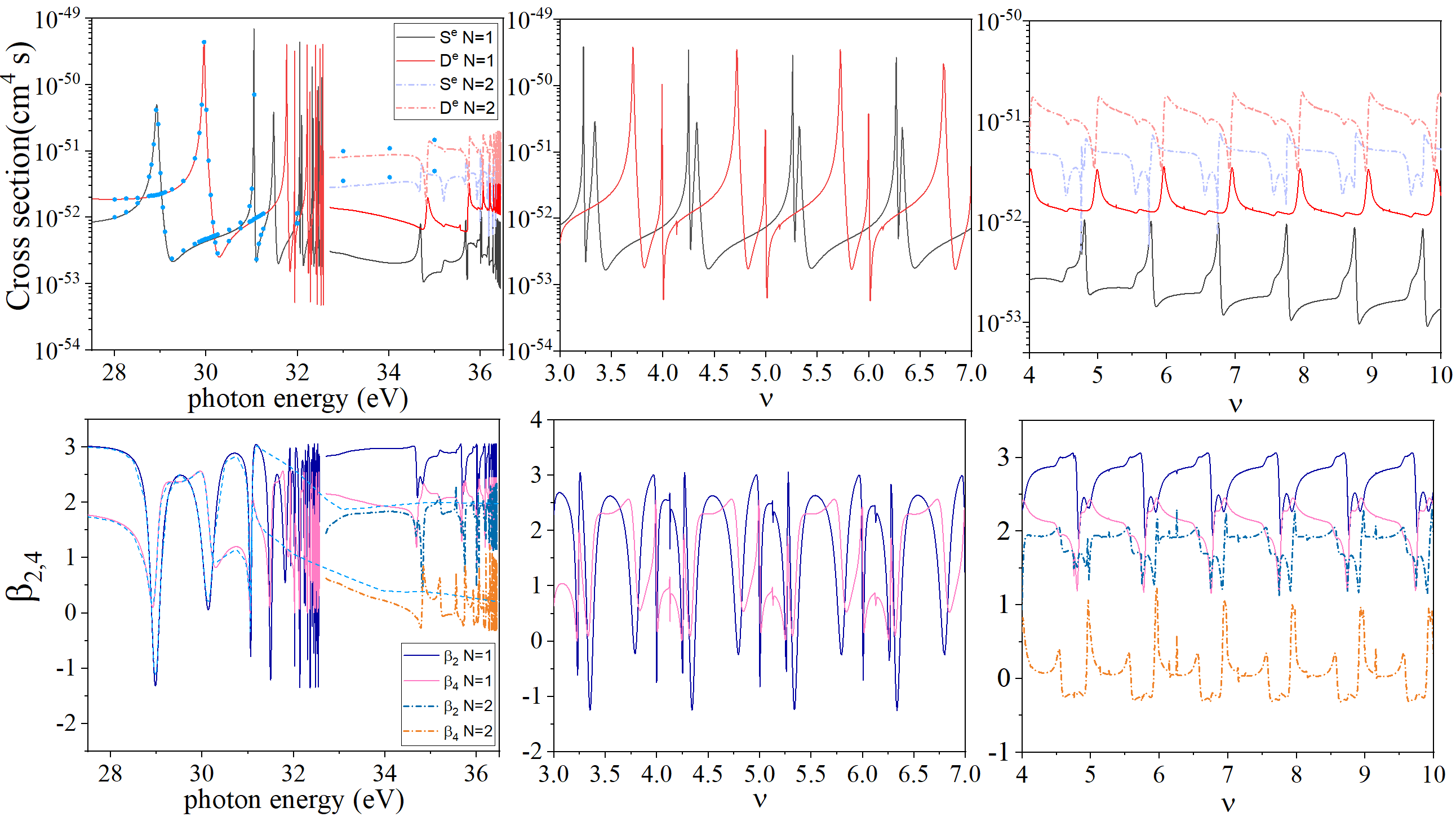}
  \caption{Partial two-photon ionization cross sections and angular distribution $\beta_{2,4}$ parameters. The left panel is calculated for final state energies in the range from -0.9 to -0.223 a.u., and plotted versus the photon energy.  The meaning of each curve is written in the inset label. The blue points are taken from Boll {\it et al.} \cite{Diego:2019}, as are the blue dashed curves for $\beta_{2,4}$.Their calculation are based on a 2 fs pulse duration so the narrow resonances are not resolved. 
  The middle two panels are an expanded version of the partial cross sections and $\beta$ parameters near the $N=2$ threshold, shown on the effective quantum number scale for the outer electron,$\nu=1/\sqrt{2(E_{N=2}-E_f)}$. The right two panels are expanded near the $N=3$ threshold, shown versus $\nu=1/\sqrt{2(E_{N=3}-E_f)}$. \textcolor{black}{To see the details of resonance structures near the $N=3$ threshold, go to Fig.~\ref{14}.}}
  \label{2}
\end{figure*}

Our first calculation is for ionization with two linearly polarized photons along a common $z$-axis, where the final state energies considered are in the range from -0.9 to -0.223 a.u. 
The generalized cross sections for reaching the $^1S^e$ and $^1D^e$ symmetries and the asymmetry parameters $\beta_2,\beta_4$ are shown in Fig.~\ref{2}.
The on-shell intermediate states for this range of energies are above the He$^+$ $N=1$ threshold at -2 a.u. so this calculation is always in the range of above-threshold ionization. The ionization threshold for $N=2$ is at -0.5 a.u. and for $N=3$ it is near -0.222 a.u. 
The full set of Rydberg states converging to the $N=2, 3$ thresholds emerge and are shown in detail in the right two graphs, displayed versus the effective quantum number \textcolor{black}{$\nu=1/\sqrt{2(E_{N}-E_f)}$} of the outer electron. 
The resonances below and above the $N=2$ threshold have opposite Fano lineshape asymmetries: The cross-sections display Fano lineshapes with negative shape ($q$) parameters in the middle graph while positive $q$ values occur in the right graph.
The resonance peaks in the calculation are associated with quasi-bound autoionizing final state levels. Our results for the first few autoionizing states are based on an R-matrix calculation that includes all the electron correlations. Table 1 compares our results with some previous results in the literature \cite{Ho:1983,Sanchez:1995}, showing good general agreement.

In addition, our results (solid lines) are compared with the Boll {\it et al.}\cite{Diego:2019} calculation that is based on a fully correlated time-dependent solution (the blue points and dashed lines) using a 2 fs laser pulse. When above the He$^+$ $N=2$ threshold, their calculation gives the total cross sections and asymmetric parameters while our calculation separate them into a fast($N=1$) and slow($N=2$) channel. 
The agreement is quite good at those energy points irrespective of the narrow resonance structures which are not resolved in their system. 
Our calculations are for monochromatic photons and include more energy mesh points, and hence are not limited in their energy bandwidth. 
Interestingly, it can be seen that the minimum of the cross-sections near the first few resonances fail to go to zero, as a feature of the above-threshold ionization.  When resonances in that region are reached via single-photon ionization, they are guaranteed to have an exact zero in an $LS$-coupling calculation because they decay into only one channel and at one energy there will be perfect destructive interference. 

\begin{table}[htbp]
\centering
\caption{The energy($E$) and and decay width($\Gamma$) for first three doubly-excited levels lying just below the $N=2$ threshold for S and D wave singlet state He determined by previous works(given by the superscripts) and our calculations.(-x) indicates $10^{-x}$. All the unit is in atomic unit(a.u.)}
\begin{tabular}{c||c||c||c}
\hline
Levels & $_N(K,T)_n$ & This work: $E(\Gamma)$  &Previous works: $E(\Gamma)$ \\
\hline
$^1S^e$(1) & $_2(1,0)_2$&-0.7780(4.70(-3))&-0.7779(4.54(-3)) {\cite{Ho:1983}} \\
$^1D^e$(1) & $_2(1,0)_2$&-0.7018(2.37(-3))&-0.7004(2.59(-3)) {\cite{Sanchez:1995}} \\
$^1S^e$(2) & $_2(-1,0)_2$&-0.6222(2.36(-4))&-0.6219(2.16(-4)){\cite{Ho:1983}} \\
$^1S^e$(3) & $_2(1,0)_3$&-0.5902(1.42(-3))&-0.5899(1.35(-3)) {\cite{Ho:1983}} \\
$^1D^e$(2) & $_2(1,0)_3$&-0.5691(5.77(-4))&-0.5687(6.17(-4)) {\cite{Sanchez:1995}} \\
$^1D^e$(3) & $_2(0,1)_3$&-0.5563(3.17(-5))&-0.5563(2.12(-5)) {\cite{Sanchez:1995}} \\
\hline
\end{tabular}
\end{table}

The next situation to be discussed is the ionization with two different laser sources with frequencies $\omega_1=2\omega_2$ and opposite circularly polarized photons. 
This is the so-called `` trefoil field" because its total electric field traces a trefoil figure, widely used by experimentalists to generate high-harmonic laser sources with arbitrary polarization \cite{pic2}. 
With two opposite circularly polarized photons, a ``parity unfavored"  $^1P^e$ symmetry, also sometimes referred to in the literature as ``unnatural parity'' for which $\pi (-1)^L=-1$, can be obtained. The photoionization pathway can be find in Fig.~\ref{1}(b). 
Its transition from intermediate to the final state $^1P^o\rightarrow ^1P^e$ is from the fact that 
$\left(    \begin{matrix}
    1 & 1& 1 \\
    0 & -1 & 1
    \end{matrix}\right)\neq 0$.

\begin{figure}[htbp]
  \includegraphics[scale=0.33]{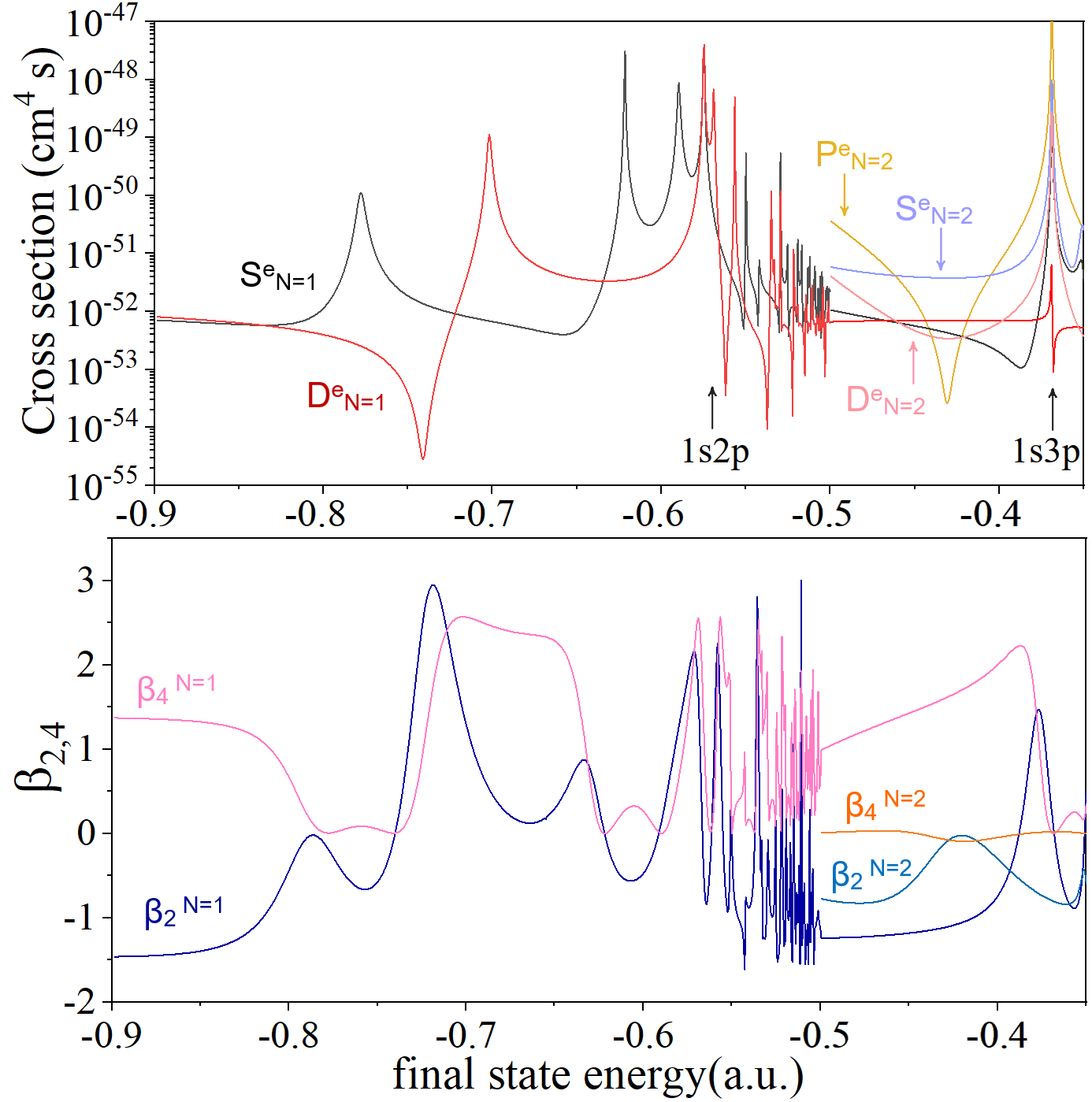}
  \caption{ Partial wave cross sections (upper) and $\beta_{2,4}$ parameters (lower) for trefoil field photoionization near the $N=2$ threshold. The parity unfavored ionization channel is present when the energy is above $N=2$ threshold at -0.5 a.u.  The two identified peaks are from the $1s2p$ and $1s3p$ intermediate state resonances.
  }
  \label{5}
\end{figure}

The cross-section formula for this situation becomes, 
\begin{equation}
\begin{split}
   \sigma_{tot}&=8\pi^3\alpha^2\omega_1\omega_2\sum_{f} \left|T_{0,f}\right|^2  \\
   T_{0,f}& =\SumInt_{\zeta} d{E_\zeta} \left(\frac{\langle f|D_2| {\zeta} \rangle
    \langle {\zeta} |D_1| 0\rangle}{E_0+\omega_1-E_{\zeta}}  + \frac{\langle f|D_1| {\zeta} \rangle
    \langle {\zeta} |D_2| 0\rangle}{E_0+\omega_2-E_{\zeta}} \right). \\
\end{split}
\end{equation}
\textcolor{black}{The dipole operator $D_i=\hat{\mathbf{\epsilon}_i} \cdot (\vec{r}_1+\vec{r}_2)$ describes the absorption of two laser photons with frequency $\omega_i$ in second-order perturbation theory. }
The relative phase and strength of the two laser field is irrelevant and factors out of the definition of the generalized cross section. Eqs.\eqref{eq2-4},\eqref{eq2-2} for the differential cross section and angular distribution asymmetry parameters still apply for this situation, since a left- and a right-handed circularly polarized light will reach $M_f=0$ final states.
The results for the partial cross-sections and asymmetry parameters are shown in Fig.~\ref{5}.

The final state energy in Fig.~\ref{5} ranges from -0.9 a.u. to -0.35 a.u. relative to the double ionization threshold, a range that reaches just across the $N=2$ threshold. 
Two intermediate state resonances are presented in the figures at -0.57 and -0.36 a.u.,corresponding to the $1s2p$ and $1s3p$ $^1P^o$ intermediate excited states. At these energies, all the final state partial-waves experience that intermediate state resonance. The other resonances are from either $^1S^e$ or $^1D^e$ final states, and they are in the same position as in Fig.~\ref{2}, giving the energy levels for double excited states. 
The parity-unfavored $^1P^e$ state in the continuum emerges only when the energy reaches the $N=2$ threshold at -0.5 a.u., since
its lowest open channel is 2p$\epsilon$p feature, so below the $N=2$ threshold there are no scattering solutions for this symmetry, although there are metastable bound states that are perfectly stable and would show up as Dirac delta functions if we were to include them.  The metastable $^1P^e$ states below the $N=2$ threshold have not been included here although they could be observed experimentally. At all energies, the non-zero asymmetry parameters always include only $\beta_{2,4}$, as is expected from Eq.\eqref{eq2-2}.

\subsection{Resonance classification}\label{rs2}

This subsection discusses the classification of the resonances and the propensity rules for their excitation and decay.
The discussion of decay propensity rules is non-trivial only for resonances above the $N=2$ threshold, since there is only one open channel($N=1$) for the electron to escape into when the final state energy is below the He$^+$ $N=2$ threshold. When more than one decay continuum exists, the relative partial decay rates of each resonance into the various alternative continuum channels can be explored, which reveals key aspects of the electron correlation physics.

Our analysis considers ionization by two identical photons that reach final state energies above the $N=2$ threshold, where in order to get maximum resolution of some very narrow resonances, the energy range analyzed here includes just one cycle of the Rydberg series converging to the $N=3$ threshold, in the range of effective quantum number, $\nu= 12-13$.

In an earlier study\cite{Petersen1991}, single-photon-absorption processes up to the He$^+$ $N=7$ threshold were measured. A ``propensity rule" was observed, indicating that the dominant channel in the excitation process should satisfy the selection rule that $\Delta A=0$, $\Delta T=1$ and $K=K_{max}$\cite{lin1986AMOP,Petersen1991}. 
For example, in the ground state helium $_1(0,0)^+$ single photoabsorption process, the $_N(N-2,1)^+$ channels play a major role . 
Another propensity rule that has been widely discussed describes the predominant decay channel of resonances. Usually the vibrational quantum number $v=\frac{1}{2}(N-1-K-T)$ is used when discussing the decaying processes. $v=0$ indicates the electrons have only a zero-point motion in the bending coordinate $\theta_{12}$.
The propensity rule for the dominant decay channel of a resonance is the following: $\Delta N=-1,\Delta A=0$, $\Delta v =0$ \cite{SG1990,SGC1992}; this rule is believed to describe the the lowest ($v=0$) and second lowest ($v=1$) channel resonances for helium atoms\cite{Sadeghpour1991}.

 The time-delay matrix is used to analysis the decay processes and the life time: $\underline{Q}^{phys}=-i\underline{S}^{phys\dagger}\frac{d\underline{S}^{phys}}{dE}$, where $\underline{S}^{phys}$ is the physical scattering matrix and $id/dE$ is the quantum time operator.
The largest eigenvalue $q_{max}$ of the Hermitian $\underline{Q}^{phys}$ gives the longest scattering delay in general and peaks whenever there is a resonance with a peak value  related to the resonance decay lifetime.
The dominant decay channel of that resonance is given by the  eigenvector corresponding to that $q_{max}$.

\begin{figure}[htbp]
  \includegraphics[scale=0.28]{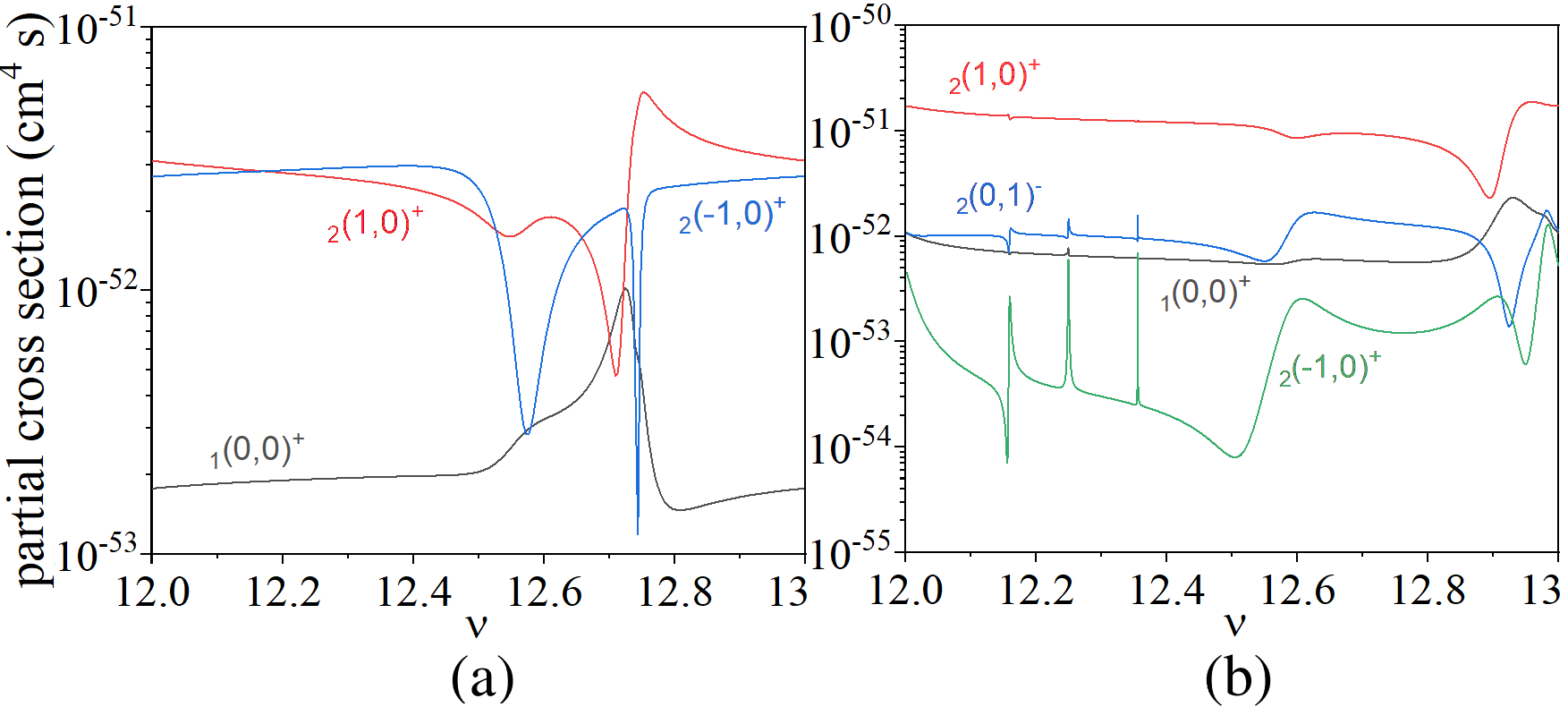}
  \caption{The two-photon ionization partial cross sections for (a) $^1S^e$ symmetry and (b) $^1D^e$ symmetry are shown for each open channel.  The effective quantum number $\nu$ is relative to $N=3$ He$^+$ threshold. 
  The $^1S^e$ excited state resonances are found to not obey the same photoabsorption propensity rule that was previously found to hold rather accurately for one-photon absorption processes that reach the $^1P^o$ symmetry.}
  \label{14}
\end{figure}

The partial cross-section for each channel is shown in Fig.~\ref{14}. The three open channels in $^1S^e$ symmetry are from the combinations of $1sns$, $2sns$ and $2pnp$, shown Fig.~\ref{14}.a. We can see the $N=2$ channels play a more important role than the $N=1$ channel, which agrees with the propensity rule that $\Delta N=-1$ primarily.
The selection rules $\Delta A=0$, $\Delta T=1$ are observed here to hold for the $^1S^e$ symmetry, but the $K=K_{max}$ rule is not satisfied according to Fig.~\ref{14}(a).
It is hard to call the $_2(1,0)^+$ channel a ``dominant channel", since it is almost the same order of amplitude comparing to $_2(-1,0)^+$ channel and even smaller at some energies. 

The four open $^1D^e$ channels, $_1(0,0)^+$, $_2(1,0)^+$, $_2(0,1)^-$ and $_2(-1,0)^+$ are from the combinations of $1snd$, $2snd$, $2pnp$, and $2pnf$, shown in Fig.~\ref{14}.b. Since the $_2(1,0)^+$ channel is dominant, the propensity rule that for each absorption process $\Delta A=0$, $\Delta T=1$ and $K=K_{max}$ are all satisfied.
Moreover, the partial cross-section decreases with the $K$ quantum number for $N=2$ channels. This is expected since the largest $K$ channel has the deepest potential at small hyperradius $R$ (shown in Fig.~\ref{13}), so both the electrons have the largest overlap with the nucleus and therefore be the most probable channel to decaying. 

\begin{figure}[htbp]
  \includegraphics[scale=0.315]{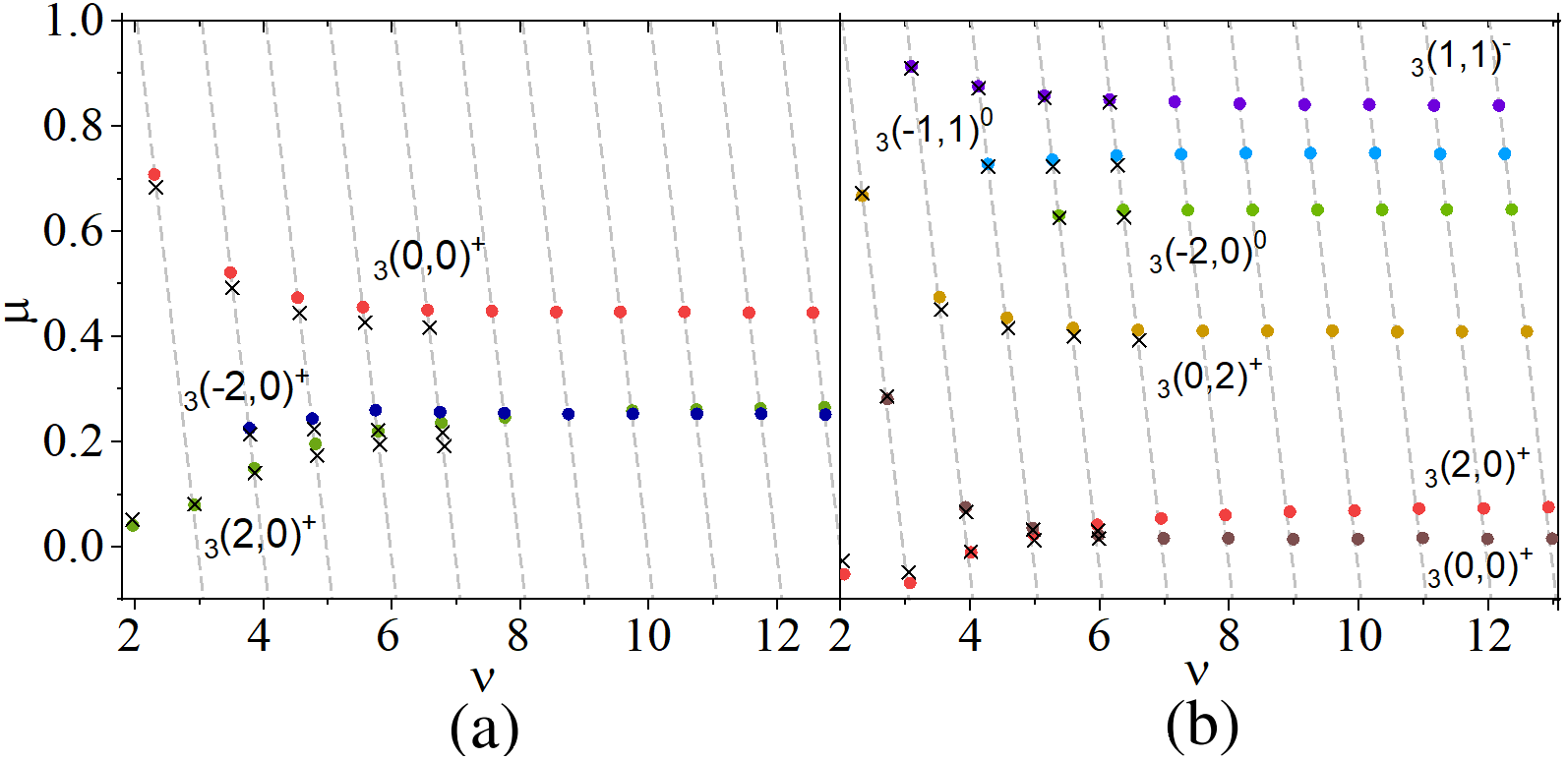}
  \caption{The quantum defect $\mu$ below $N=3$ threshold is plotted versus the effective quantum number $\nu$ for $^1S^e$(left) and $^1D^e$(right) wave bound states. The black crosses are the truncated diagonalization calculation from \cite{LAC1977}. The grey dashed curve has slope -1 and each bound state must lie on one of them because of the QDT condition $\sin{\pi(\mu+\nu)}=0$. }
  \label{17}
\end{figure}

An interesting feature in both $^1S^e$ and $^1D^e$ lineshapes is the overlap of two different resonances, i.e. the $_3(2,0)^+$ and $_3(-2,0)^+$ resonances for $^1S^e$ and $_3(2,0)^+$ and $_3(0,0)^+$ resonances for $^1D^e$. This occurs not only at high $\nu$, as from $\nu>4$ the two resonances are already very close, and at some point they overlap, as is shown in Fig.\ref{17}.

\begin{figure}[htbp]
  \includegraphics[scale=0.28]{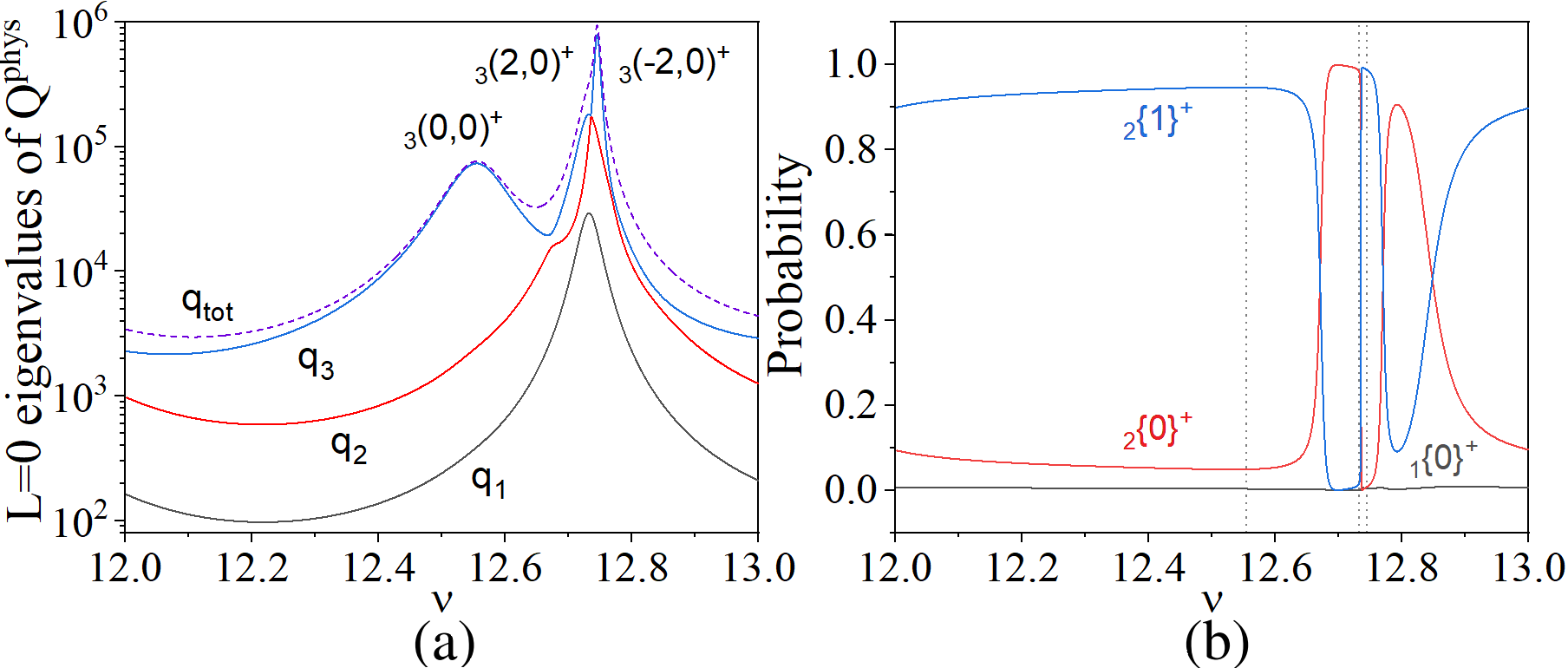}
  \caption{Time delay analysis for $^1S^e$. The left figure shows eigenvalues of the the time-delay matrix and their sum $q_{tot}$, which gives the decay lifetime, with the peaks of the curves corresponding to the position of resonances, whose character are labeled.
  The right figure shows decay probabilities to each ionization continuum channel. The dashed lines gives the position for the peaks of $q_{tot}$. The $_N\{v\}^A$ quantum numbers of the resonances are from left to right, $_3\{1\}^+$,$_3\{0\}^+$, and $_3\{2\}^+$. 
  }
  \label{15}
\end{figure}

Next, consider the decay process for the autoionization states based on the time-delay matrix analysis.
For $^1S^e$ symmetry, there are three closed channels that contribute the three resonances in each cycle of $\nu$. The resonance channel correspondences are shown in Fig.~\ref{15}(a). All three resonances are broad ones since they all have $A=+1$.
Fig.~\ref{15}(a) plots eigenvalues $q$ of the the time-delay matrix which conveys the positions and the decay lifetimes, and (b) plots the decay probability into each open channel.
Three open channels can serve as decay routes, namely  $_1(0,0)^+$, $_2(1,0)^+$ and $_2(-1,0)^+$, where the vibrational quantum number $v$ is used to delineate those channels. 
The first thing to note is that the decay probability into the $N=1$ channel is almost negligible, obeying the propensity rule that $\Delta N=-1$.
For $_N\{v\}^A=_3\{0\}^+$ and $_3\{1\}^+$ resonances, the propensity rule $\Delta A=0$, $\Delta v =0$ is satisfied. For the $_3\{2\}^+$ resonance, there is no $\Delta v=0$ continuum, so it decays to the $_2\{1\}^+$ channel that has the minimum $|\Delta v|$.

\begin{figure}[htbp]
  \includegraphics[scale=0.28]{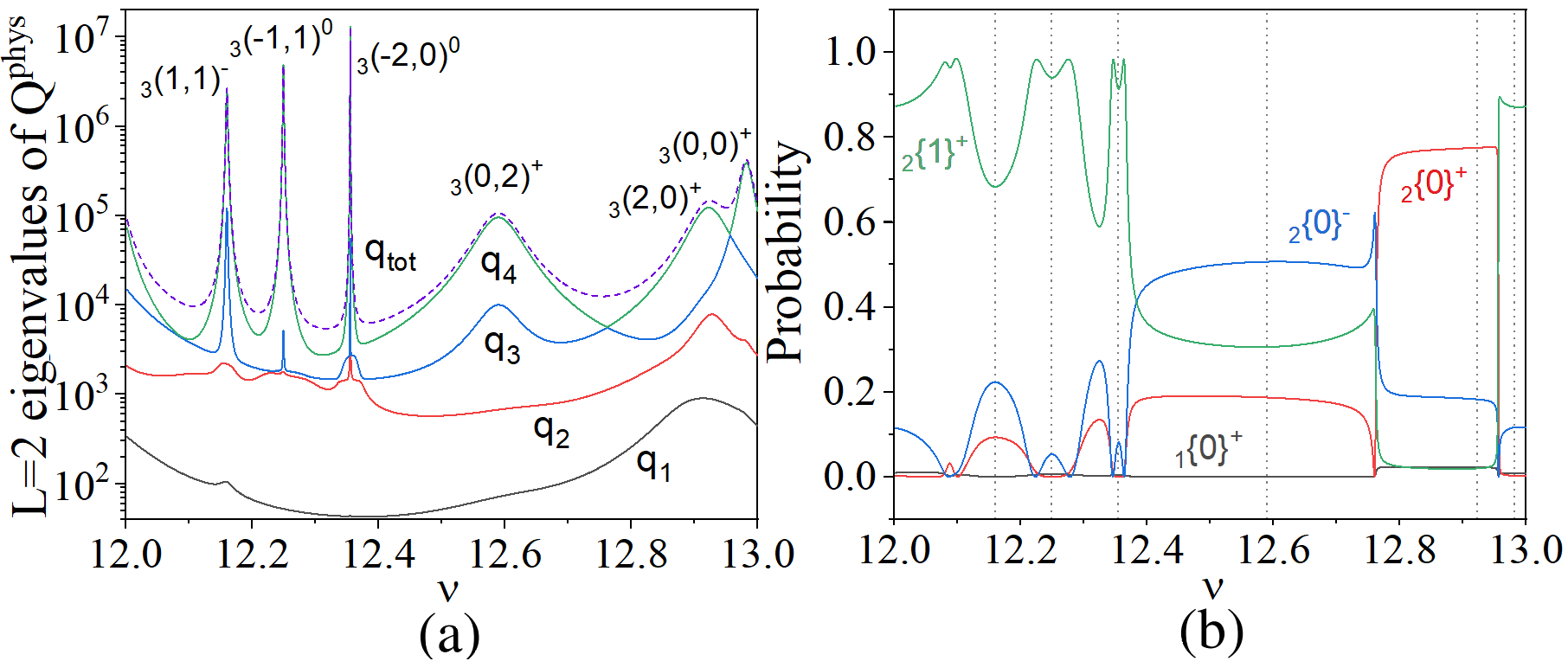}
  \caption{Time delay analysis for $^1D^e$ symmetry resonances. The left figure shows the eigenvalues of the the time-delay matrix and the right figure shows partial decay probabilities. The labels and notation are the same as Fig.~\ref{15}. The $_N\{v\}^A$ quantum numbers of the resonances are from left to right, $_3\{0\}^-$,$_3\{1\}^0$, $_3\{2\}^0$,$_3\{0\}^+$,$_3\{0\}^+$and $_3\{1\}^+$.}
  \label{16}
\end{figure}

For $^1D^e$ symmetry, there are six closed channels in the range below the $N=3$ threshold, which gives six resonances in each cycle of $\nu$. Their resonance channel correspondence is given in Fig.~\ref{16}(a) The resonances with $A=+$ are broad ones and the others are narrow, as expected from their potential curves. Their decay probabilities into four open channels $_1(0,0)^+,_2(1,0)^+, _2(0,1)^-$ and $_2(-1,0)^+$ are shown in Fig.~\ref{16}(b).
The probability of decaying into the $N=1$ channel is again negligible. However, the other propensity rules are not that obviously satisfied.
The propensity rule is observed here to hold only for the most prominent resonance $_3(2,0)^+(_3\{0\}^+)$ and the third prominent resonance $_3(0,0)^+$. For the second prominent resonance $_3(0,2)^+(_3\{0\}^+)$, its decay probabilities are 0.51 to $_2\{0\}^-$, 0.31 to $_2\{1\}^+$ and 0.18 to $_2\{0\}^+$, not consistent with the expected propensity rule. 
It is noted that in this situation both $_3(2,0)^+$ and $_3(0,2)^+$ have been denoted as $(_3\{0\}^+)$ channel. Their difference is their $T$ value, and we note that $T=2$ resonance decays have apparently not been discussed in earlier work.
The prominent channel for all three narrow resonances is $_2\{1\}^+$. Those narrow resonances are hardly observed in experiments and are not necessarily expected to be governed by the propensity rule.

To test the validity of our calculations, we plot the quantum defect versus the effective quantum number for the $^1S^e$ and $^1D^e$ symmetry, from the first bound state above $N=2$ threshold up to $\nu<13$,  as shown in Fig.~\ref{17}.
The black crosses are the calculation results from \cite{LAC1977}, Table VII, based on the truncated diagonalization method. Some calculations from more recent papers\cite{Eiglsperger,Ho:1985} are given and compared in Table II. For the $^1D^e$ wave, the differences in $\nu$ are around 0.01 and the agreement is quite good, but for the $^1S^e$ wave the differences are slightly larger, especially as $\nu$ increases.
The grey dashed lines have slope -1 and connect all the bound states within each $\nu$ circle, since the bound state energies satisfy $\sin{\pi(\mu+\nu)}=0$. 
Evidently, the $^1S^e$ symmetry has all the resonances series present above  $\nu\geq 3$, and the $^1D^e$ wave from $\nu\geq5$, since the hyperspherical potentials for some of the channels are rather high lying and repulsive at small hyperradii.
In addition, there is a cross-over in the $^1S^e$ wave $_3(2,0)^+$ and $_3(-2,0)^+$ and the $^1D^e$ wave $_3(2,0)^+$ and $_3(0,0)^+$ series, indicating that the those resonances are very close to each other and even overlap when those curves across.

\begin{table}[htbp]
\centering
\caption{The energy(E) and and decay width($\Gamma$) for first three doubly-excited levels lying just below the $N=3$ threshold for S and D wave singlet state He determined by previous works(given by the superscripts) and our calculations.(-x) indicates $10^{-x}$. All quantities are in atomic units(a.u.)}
\begin{tabular}{c||c||c||c}
\hline
Levels & $_N(K,T)_n$ & This work: $E(\Gamma)$  &Previous works: $E(\Gamma)$  \\
\hline
$^1S^e$(1) & $_3(2,0)_3$&-0.3534(3.00(-3))&-0.3535(3.01(-3)) {\cite{Eiglsperger}} \\
$^1D^e$(1) & $_3(2,0)_3$&-0.3430(4.74(-3))&-0.34314(5.25(-3)) {\cite{Ho:1985}} \\
$^1S^e$(2) & $_3(0,0)_3$&0.3174(6.32(-3))&-0.31745(6.66(-3)) {\cite{Eiglsperger}}\\
$^1D^e$(2) & $_3(0,2)_3$&-0.3154(4.07(-3))&-0.3157(4.30(-3)) {\cite{Ho:1985}}\\
$^1D^e$(3) & $_3(0,0)_3$&-0.2900(1.41(-3))&-0.2900(1.20(-3)) {\cite{Ho:1985}} \\
$^1S^e$(3) & $_3(2,0)_4$&-0.2809(1.65(-3))&-0.2811(1.502(-3)) {\cite{Eiglsperger}} \\
\hline
\end{tabular}
\end{table}

\section{summary }\label{end}

In summary, we have studied the autoionizing states and the angular distribution of photoelectrons for two-photon ionization of helium atoms, with an exploration of nontrivial, nonperturbative electron correlation patterns.
The partial cross-sections and asymmetry parameters are computed in the energy range associated with the simplest example of non-sequential above-threshold ionization.
The two-identical-photon ionization has been computed up to the He$^+$ $N=3$ threshold, and the lineshapes, angular distributions, including the autoionizing state positions and decay processes.  They have also been compared to some earlier theoretical studies, showing generally good agreement.
The two-photon ionization has also been computer for an incident trefoil field, which generates a parity-unfavored ionization channel.
Our calculation shows that for the above-threshold ionization, the minimum of the cross-section at final state Fano lineshapes does not always go to zero. A detailed structural analysis of the Rydberg series and their decays has also been presented.
We also compute two-photon ionization up to the He$^+$ $N=3$ threshold and discussed the propensity rule for the two-photon cases with our results.
For the photoabsorption processes near the He$^+$ $N=3$ threshold, one still observes propensity for exciting $\Delta A=0$, $\Delta T=1$, but the excitation rule $K=K_{max}$ is not always satisfied. For autoionization decay, the propensity rule $\Delta N=-1,\Delta A=0,\Delta v =0$ is satisfied only for the most prominent resonances, and for the second prominent resonances in $^1D^e$ wave when $T=2$, no clearly dominant decay channel exists.
One possible extension to this work would be to investigate the alignment and orientation of excited helium ions produced in the two-photon ionization process, where the ionic electron alignment, for example, was found to mimic the behavior of the photoelectron angular distribution in a single-photon ionization process\cite{PhysRevA.39.115}.

\begin{acknowledgments}
  This material is based upon work supported by the U.S. Department of Energy, Office of Science, Office of Basic Energy Sciences, under Award Number DE-SC0010545.
\end{acknowledgments}

\bibliographystyle{unsrt}
\bibliography{He_Multiphoton}

\end{document}